\documentclass[final,5p,times,twocolumn]{elsarticle}
\usepackage{lineno}
\usepackage{setspace}

\usepackage{amssymb}
\usepackage{float}
\usepackage{ulem}
\usepackage[pdftex]{color}

\begin{document}

\begin{frontmatter}



\title{Grain-size dependence of water retention in a model aggregated soil}


\author[N1]{Hyuga Yasuda}
\author[N1]{Makoto Katsura}
\author[N1]{Hiroaki Katsuragi}
  
  \address[N1]{Department of Earth and Space Science, Osaka University, 1-1 Machikaneyama, Toyonaka 560-0043, Japan}

  
\begin{abstract}
We experimentally examined the amount of water retention in a model soil composed of aggregated glass beads. The model soil was characterized by two size parameters: size of aggregates $D$ and size of monomer particles (composing aggregates) $d$. In the experiment, water was sprinkled on the model-soil system that has an open top surface and drainable sieve bottom. When the sprinkled water amount exceeded a threshold (retainable limit), draining flux balanced with the sprinkled flux.
The weight variations of retained and drained water were measured to confirm this balanced (steady) state and quantify the retained water. 
We defined the weight of the retained water in this steady state as $W_0$ and examined the relationship among $W_0$, $d$ and $D$. As a result, it was revealed that $W_0$ increases as $d$ decreases simply due to the capillary effects. Regarding $D$ dependence, it turned out that $W_0$ becomes the maximum around $D\simeq 500$~$\mu$m. The value of $D$ maximizing water retention is determined by the void formation due to the aggregated structure, capillary effect, and gravity.
\end{abstract}



\end{frontmatter}


\section{Introduction}
Water retention in porous media has been studied in various fields. Glass beads are frequently used as a model material to simplify the structure of porous media in which water can be retained~\cite{HERMINGHAUS1,HERMINGHAUS2}. One of the most significant advantages of using glass beads is the spherical shape that enables us to simply analyze grains contact network and pore structures. Therefore, spherical grains have been used for various experiments~\cite{Geistlinger:2015,Vincens:2015,LV:2022}.
Moreover, most of the numerical works have used sperical grains~\cite{Gao:2012,YANG:2013}. Thus, to compare the experimental results with the numerical ones, spherical grains are better. Even by using spherical grains, physical properties of the mixture of grains and water (wet granular matter) complexly depend on the wetting degree~\cite{Schubert:1977,Mitarai:2006,Herminghaus:2013}. 

\begin{figure*}
  \centering
  \includegraphics[width=16cm]{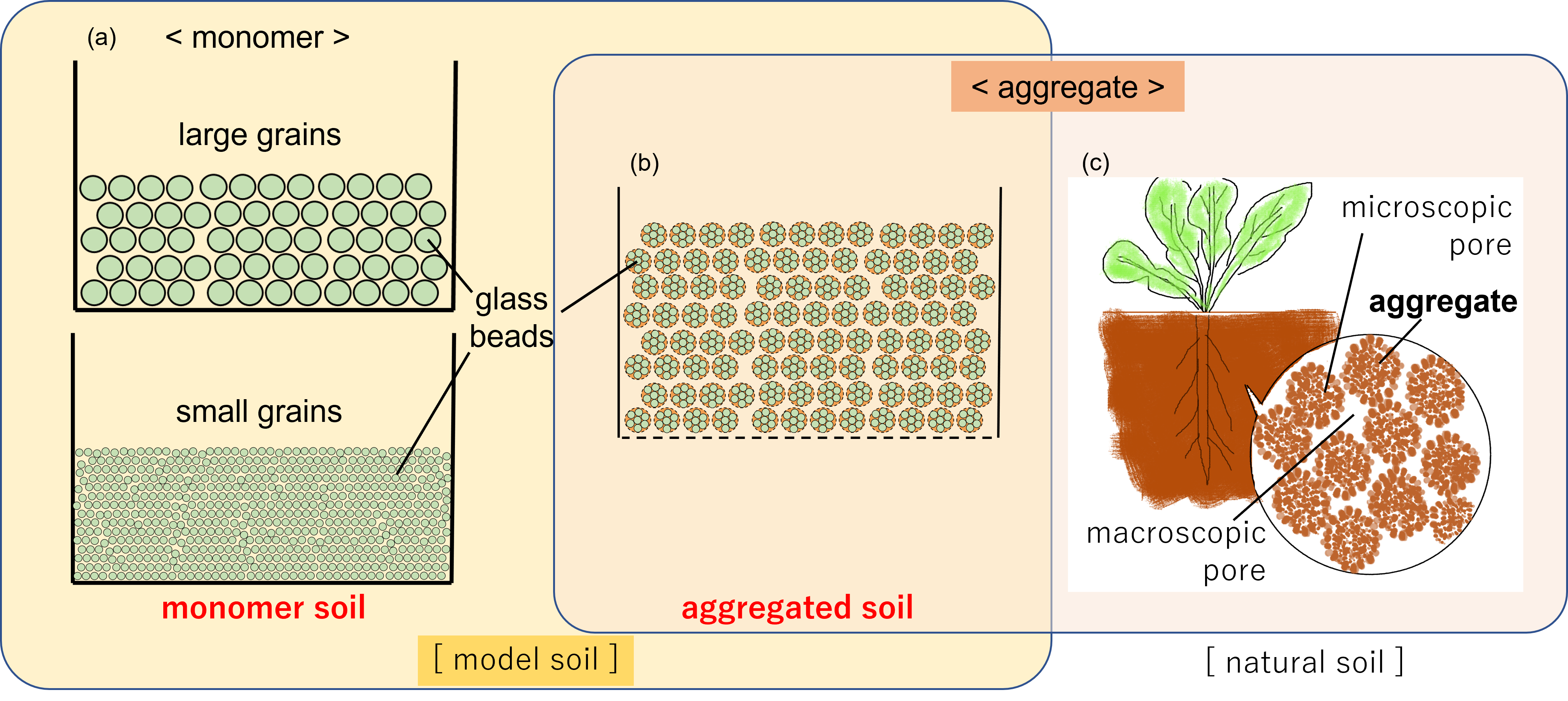}
  \caption{Schematic diagrams of structures of (a) monomer soil, (b) aggregated soil, and (c) natural soil. In this figure, diameter of the grains shown in the upper panel of (a) is same as that of aggregates in (b). Diameter of the grains shown in the bottom panel of (a) is same as that of monomer particle size composing aggregates in (b).}
  \label{structures}
\end{figure*}

Monodisperse glass beads are certainly too simple to model natural porous materials. 
	For example, natural soil has hierarchical structures since tiny particles such as clay grains often form aggregates using organic substances to stick together (Fig.~\ref{structures}(c)). There are two distinct pore-size scales in such soil structure. The larger pore between aggregates is called macroscopic pore and the smaller pore within each aggregate is called microscopic pore. In order to mimic such hierarchical structures, lightly sintered glass beads have been used for forming aggregates~\cite{agg1,agg2}. Sintered glass beads are useful to keep the structure stable~\cite{okubo} and estimate the pore size distribution. By using the sintered glass-beads aggregates for the model soil, the simple structure of capillary bridge between grains can also be assumed.
	
In most of the previous studies, water retention in soil has been evaluated by establishing correlations between water content and pressure head~\cite{kosugi}. Water supply by precipitation and gravity-driven drainage in soil were not directly modeled. However, considering the actual situation such as rainwater permeating into soil, the amount of retained water must be determined by the balance between the precipitation rate, drainage, and capillary suction. Besides, since the drainage process of the retained water depends on the initial water content, the entire process (from wetting to drying) should be taken into account in order to evaluate the actual water retention in soil.

As a simple experiment studying water evaporation from soil, Kondo et al.~\cite{kondo} focused on the drying phase in model soil composed of glass beads by measuring the weight variation of the sample which was initially saturated with water. However, hierarchical structures were not considered in their experiment because they used monodisperse glass beads (Fig.~\ref{structures}(a)). In addition, their experimental system could not have the drainage effect because the vessel they employed had a closed bottom base. In natural soil, the gravity-driven drainage affects the water retention and drying processes.

Therefore, in this research, we developed an experimental set-up which evaluates the evolution of the water retention as a result of wetting and drainage processes using model soil consisting of hierarchically structured aggregates (Fig.~\ref{structures}(b)). In the experiment, hierarchical structure of soil was mimicked by using the collection of sintered glass-beads aggregates. The effect of drainage was also considered by employing an open-bottom vessel. To control the water retention degree, two parameters, size of aggregates $D$ and monomer particle size $d$, were systematically varied (Fig.~\ref{sample}(a)). Such hierarchical structure significantly affects physical behaviors of dry granular materials~\cite{Katsuragi:2018,Felipe:2021}. The effect of granular hierarchy in wet granular matter has not been studied yet. Using this setup, we can evaluate the effects of hierarchical structure of soil and water drainage in wetting and drying processes. In general, wetting and drying processes of porous granular media are complex~\cite{Xu:2008,Xiong:2014,Wei:2014,Zhonghao:2019}. Here, we simply analyze the retainable water content in the hierarchically structured model soil. Although we have confirmed the drying curves that are similar to the observation in evaporation from porous media~\cite{Hongcai:2015}, the obtained data are still preliminary. The entire drying process will be discussed elsewhere in future. Thus, we focus only on the water retention in this study.

As a first step to understand the complex nature of water retention in soil, we measured the water retention of the precipitated soil. Particularly, we focus on the relation between the amount of retained water  and two size parameters $D$ and $d$. 

\section{Experiment}
\subsection{Sample preparation}
\label{porosity}
\begin{figure}
  \centering
  \includegraphics[width=8cm]{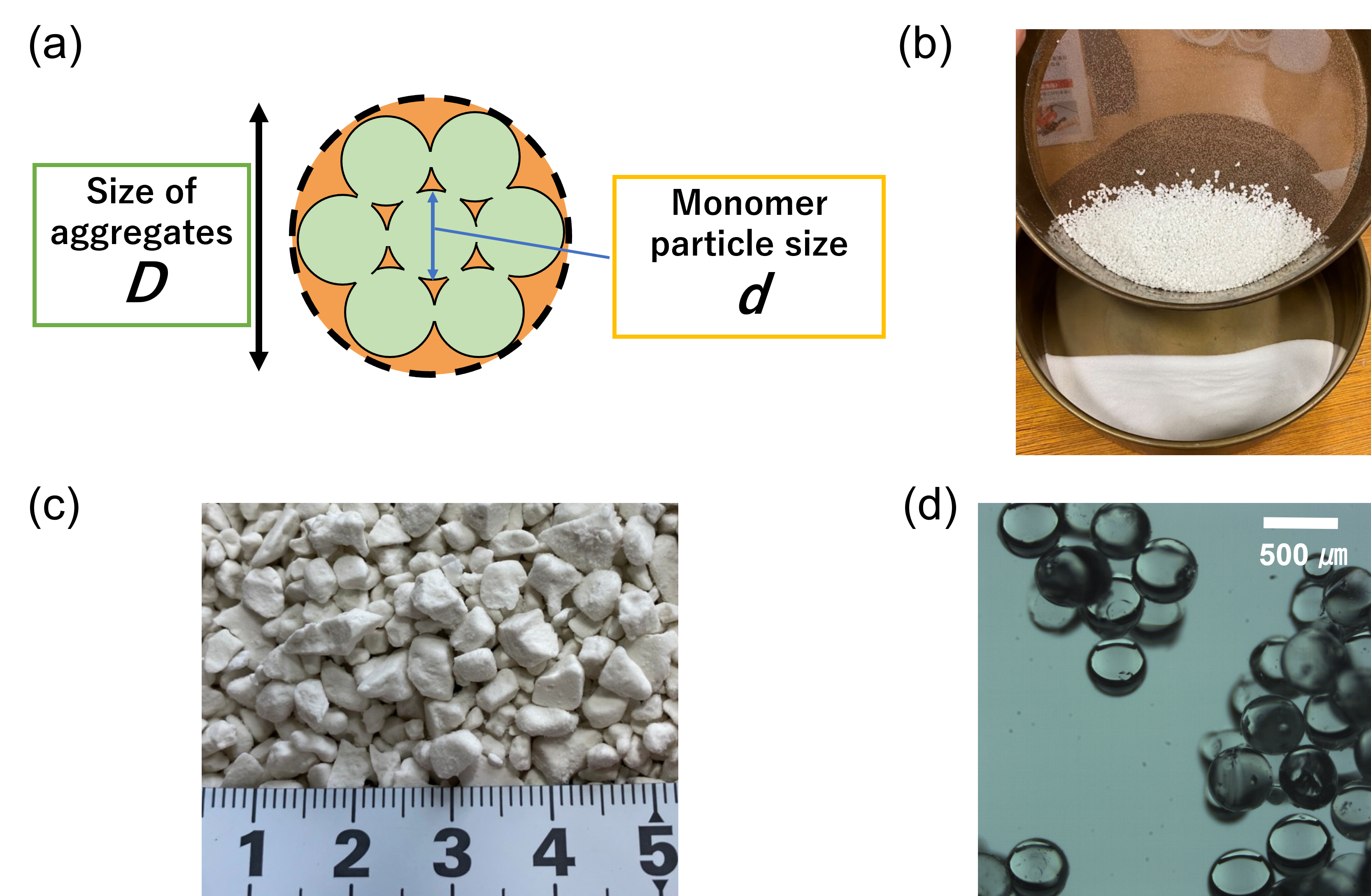}
  \caption{(a) Definitions of parameters $D$ and $d$, (b) sifting process, (c) actual photo of the aggregated glass beads ($d$ = 18 $\mu$m, $D$ = 2000--4760 $\mu$m), and (d) microscope observation of sintered glass beads ($d$ = 400 $\mu$m). The neck structure made by sintering is negligibly small.}
  \label{sample}
\end{figure}

Glass beads were employed as monomer materials to form a model soil having hierarchical structure. The representative diameters of the glass beads used in the experiment were 5, 18, 100, 400, 2000 and 3000 $\mu$m (Potters-Ballotini Co., Ltd., EMB-10, P-001, As One Corp., BZ-01, 04, 2, and N6326450010302; true density $\rho_g$ = 2.5--2.6~$\mathrm{g/cm^3}$). We prepared the hierarchically structured soil by sintering (650~\(^\circ\)C, 50--90 min) a cluster of tiny monomer glass beads ($d$ = 5, 18, 100, 400 $\mu$m). The chunk of sintered glass beads was crushed and sifted in a sieve to form aggregated grains with various size ranges (Fig.~\ref{sample}(b),(c)).
The aggregates were classified by their sizes $D$ as XS: 74--250 $\mu$m, S: 250--840 $\mu$m, M: 840--2000 $\mu$m, and L: 2000--4760 $\mu$m. 

Glass beads of $d$ = 18 $\mu$m were used as monomers for creating all of those aggregate samples (from XS to L). The corresponding packing fraction $\phi$ is $0.29$, $0.28$, $0.31$, and $0.36$, respectively (Table.\ref{packing}). The packing fraction $\phi$ was obtained from $\rho_g$ and bulk density by measuring the bulk volume and weight in a cylinder in diameter of 4.7 cm. Glass beads of $d$ = 5, 18, 100, and 400 $\mu$m were used for creating L-sized aggregates. They respectively have $\phi=0.28$, $0.36$, $0.34$, and $0.36$. The dependency of the water content on $d$ was discussed by using L-sized aggregates consisting of $d$ = 5, 18, 100 and 400 $\mu$m glass beads and $D$ dependence was investigated by using 18 $\mu$m glass beads forming various sizes of aggregates: XS, S, M and L. Non-aggregated (monomer) glass beads (5--3000 $\mu$m) were also used as monomer model soils. The packing fraction $\phi$ of 5 $\mu$m and 18 $\mu$m monomer glass beads are 0.49 (after compression) and 0.54, respectively, and the other monomer glass beads (100--3000 $\mu$m) have $\phi=0.60$. Here, the samples achieve random close packing except for 5 $\mu$m and 18 $\mu$m glass beads. Monomer glass beads of $d$ = 5 $\mu$m were compressed because the initial volume was too large to pack into the vessel. The packing fraction of $d$ = 5 and 18 $\mu$m becomes smaller than that of the random close packing. For the aggregate samples, the bulk packing fraction $\phi$ is a product of microscopic one $\phi_{\rm{micro}}$ and macroscopic one $\phi_{\rm{macro}}$, $\phi = \phi_{\rm{micro}} \times \phi_{\rm{macro}}$. We consider $\phi_{\rm{macro}}$ is close to the random close packing while $\phi_{\rm{micro}}$ depends on $d$.

\begin{figure}
  \centering
  \includegraphics[width=8cm]{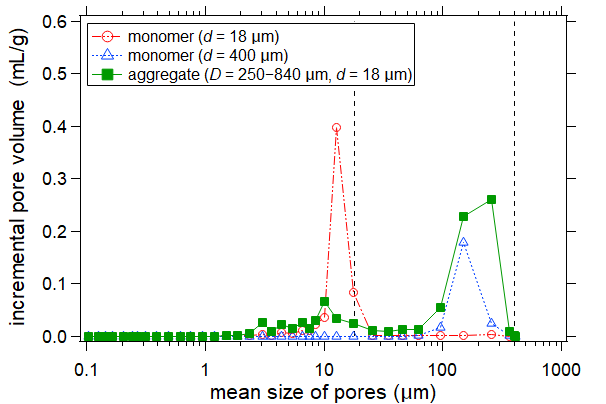}
  \caption{The incremental pore volume as a function of mean size of pores. The vertical dashed lines show the representative diameter of grains of 18 $\mu$m and 400 $\mu$m. The scale of pore size almost corresponds to grain size. The peak around 10 $\mu$m in the aggregate plot is originated from microscopic pore in the aggregate composed of 18 $\mu$m glass beads. The peak around 150--200 $\mu$m is caused by macroscopic pore between aggregates with size of $D$ = 250--840 $\mu$m.}
  \label{pore}
\end{figure}

Fig.~\ref{sample}(d) shows a microscope observation of aggregates consisting of 400 $\mu$m glass beads. Roughly the shape of the monomer beads was kept spherical in the sintered aggregates. Therefore, we can neglect the deformation of the glass beads due to the neck formation between the connecting monomer grains. The pore distribution was characterized by mercury intrusion porosimetry~\cite{okayama,Herbert:2006}. Two types of monomers ($d$ = 18 and 400 $\mu$m) and S-sized aggregates consisting of $d$ = 18 $\mu$m beads were measured as shown in Fig.~\ref{pore}. The vertical dashed lines in Fig.~\ref{pore} display the diameters of 18 $\mu$m and 400 $\mu$m. As expected, the representative size of pores almost matches the diameters of grains. Representative pore size is about two to four times smaller than the constituent grains as previously reported~\cite{Herbert:2006}. The plot of aggregate sample shows bimodal shape which originates from microscopic and macroscopic pores.

\begin{table}
 \centering
 \caption{Prepared grains and their packing fraction $\phi$}
 \label{packing}
  \begin{tabular}{cccc}
   \hline
   type & $D$ ($\mu$m) & $d$ ($\mu$m) & $\phi$ \\
   \hline \hline
   monomer & 5 & 5 & 0.49 \\
   & 18 & 18 & 0.54 \\
   & 100 & 100 & 0.60 \\
   & 400 & 400 & 0.60 \\
   & 2000 & 2000 & 0.60 \\
   & 3000 & 3000 & 0.60 \\
   \hline
   XS & 74--250 & 18 & 0.29 \\
   \hline
   S & 250--840 & 18 & 0.28 \\
   \hline
   M & 840--2000 & 18 & 0.31\\
   \hline
   L & 2000--4760 & 5 & 0.28 \\
   & 2000--4760 & 18 & 0.36 \\
   & 2000--4760 & 100 & 0.34 \\
   & 2000--4760 & 400 & 0.36 \\
   \hline
  \end{tabular}
\end{table}

\subsection{Setup}
\begin{figure}
  \centering
  \includegraphics[width=8.5cm]{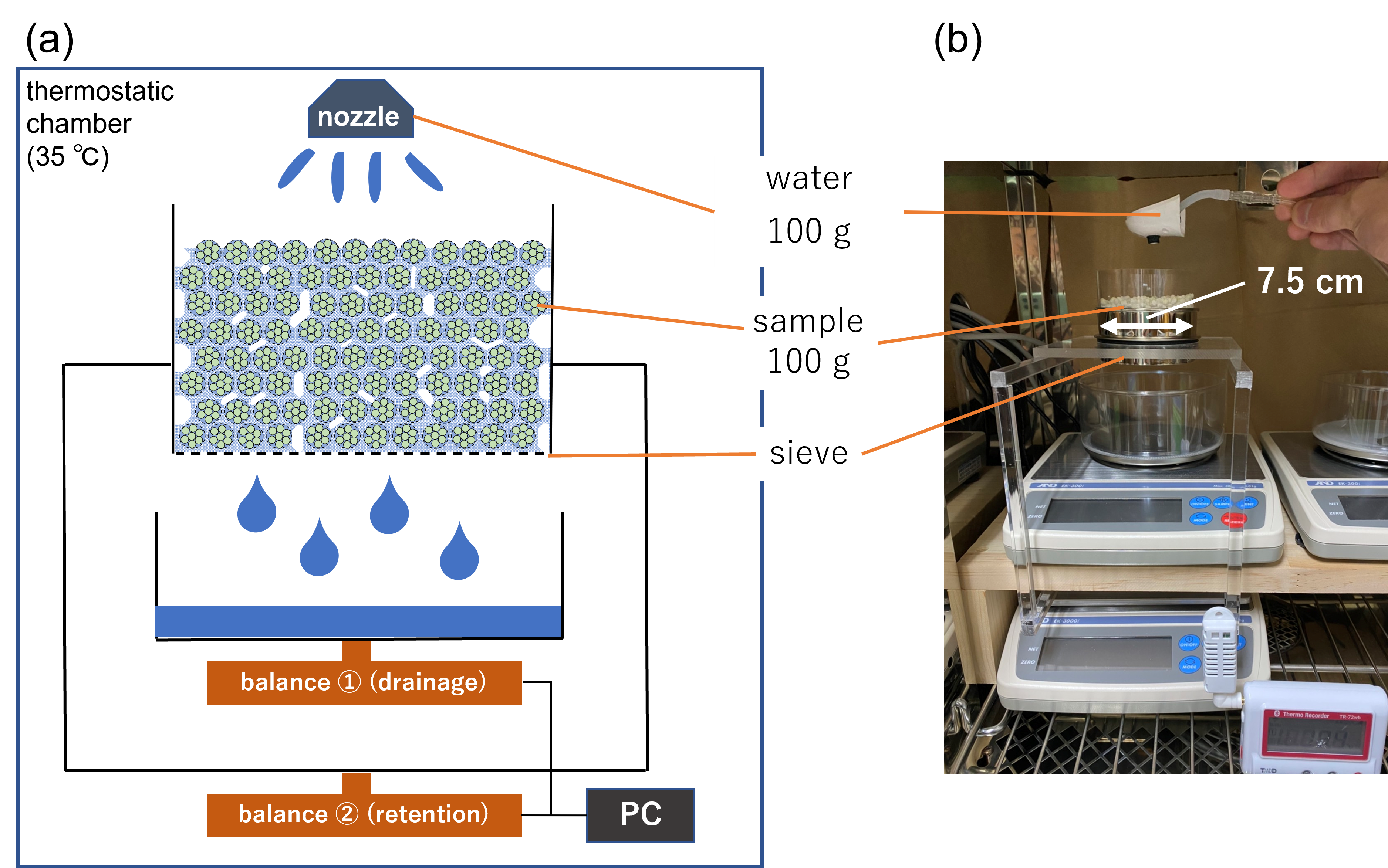}
  \caption{(a) Schematic diagram and (b) real photograph of the experimental apparatus. Masses of the drained water and the retained water (including soil mass) were measured by the electronic balance 1 and 2, respectively. All the experiments were performed in the constant temperature (35~\(^\circ\)C), and the humidity range was in the range of 16--32\%.}
  \label{setup}
\end{figure}

Aggregated dry particles of the fixed mass of 100 g were poured in a vessel whose bottom consists of a sieve in diameter of 7.5 cm with 150 $\mu$m of opening (Fig.~\ref{setup}). Typical sample thickness $H_{\rm{soil}}$ ranged 1.5--3 cm depending on $\phi$. To prevent the leakage of tiny grains, a paper filter (Whatman, Cat No 1001 090, cut into a circle in diameter of 7.5 cm) was put on the bottom sieve. Then, the fixed amount of water (100 g) was sprinkled on its surface for about 4 minutes with a flowrate 0.45 g/sec (which corresponds to 370 mm/h precipitation intensity) by using a spray nozzle (dretec SD-800). The nozzle was held by hand to spray water all over the surface. The distance between the nozzle and surface of the model soil was in the range of 3--4.5 cm depending on $H_{\rm{soil}}$. Since the sprinkled water did not deform the sample surface at all, the effect of water inertia was negligibly small.
The temporal variations of the weight of drained water and soil including retained water were measured by electronic balances (A\&D Co., Ltd., EK-300i) connected to PC (Fig.~\ref{setup}(a)). All the experiments were performed under constant temperature of $35$~$^{\circ}$C kept by the thermostatic chamber (Isuzu Seisakusho Co., Ltd., VTR-115). From the measured mass variation, we analyzed the water retention ability of the hierarchically-structured model soil. In other words, we simply measured how much water the open soil can retain under the precipitation condition. We also measured the ambient humidity in the chamber throughout the experiments (A\&D Co., Ltd., TR-72wb).

\section{Results and Discussion}
\begin{figure}
  \centering
  \includegraphics[width=8cm]{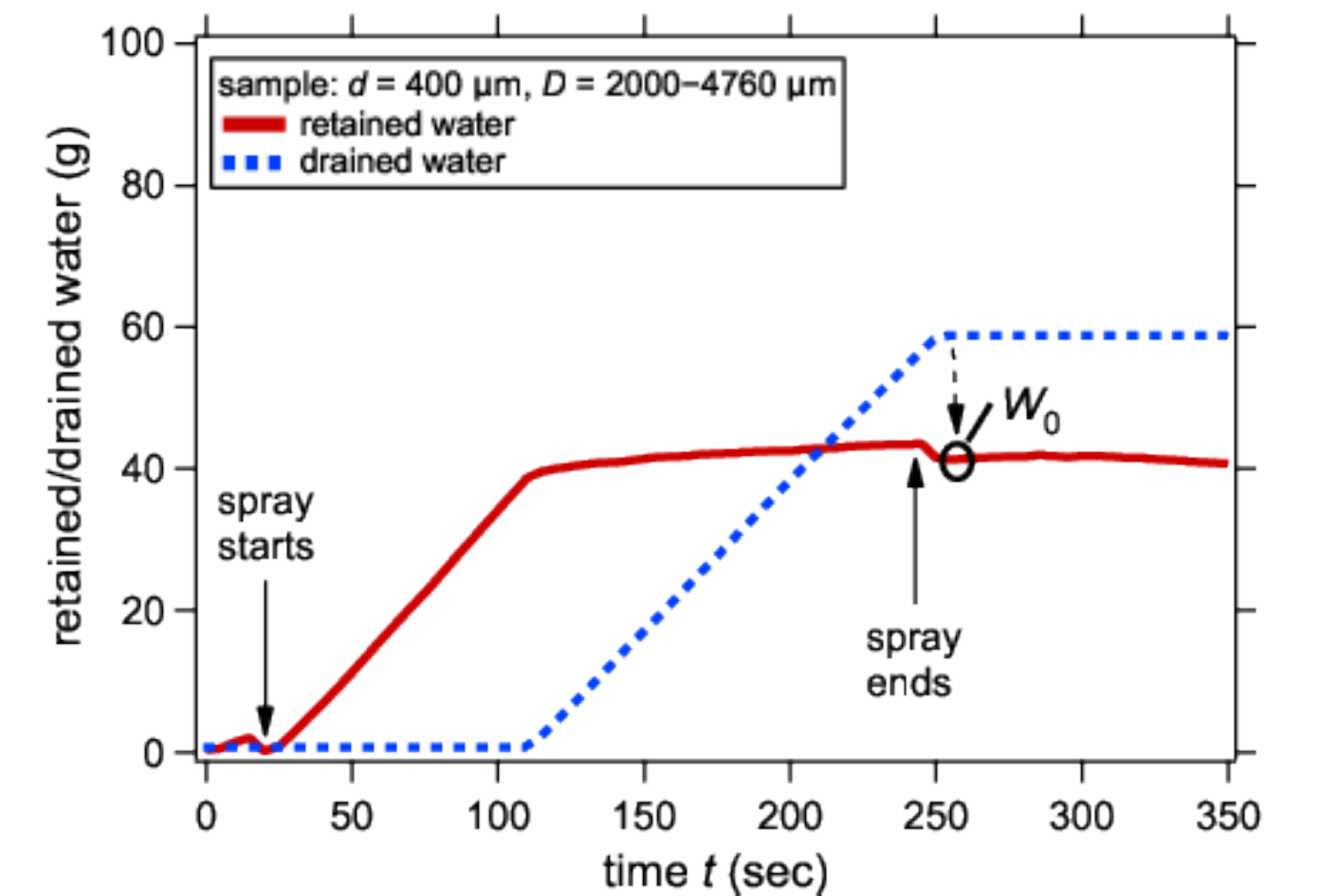}
  \caption{The amount of retained water and drained water as a function of time $t$. The retained water amount was computed by subtracting the initial (dry) mass from the measured mass of the soil sample.}
  \label{rawdata}
\end{figure}

The weights of retained water and drained water were obtained as a function of time $t$ (Fig.~\ref{rawdata}). In Fig.~\ref{rawdata}, we show the result of L-sized aggregates with $d$ = 400 $\mu$m as a representative retention/drainage graph. Three experimental runs were performed for each experimental condition to check the reproducibility. The errors shown in the following plots were calculated by standard deviation of three experimental runs. When the water supply was started, the amount of retained water began to increase. Then, within the short timescale,  the retained water became almost constant after the drainage started. In this regime, incoming water and outgoing water are balanced and reaching the steady state. Although the drainage lasted for a few seconds after stopping the water supply, both weights finally approached their asymptotic values. This tendency was common for all the experimental results. We consider that the amount of retained water $W_0$ in this state is one of the key parameters to evaluate the water retention in soil. Thus, we measured $W_0$ after the drainage flow settled (Fig.~\ref{rawdata}). Although the ambient humidity varied in the range of 16--32\%, its effect on evaporation rate is negligible in the timescale of the experiments ($\sim$ 5 min) since the total drying time was at least over 5 hours which is 60 times longer.

\begin{figure}
  \centering
  \includegraphics[width=8cm]{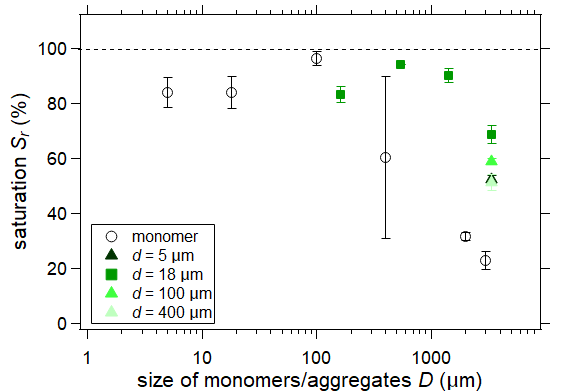}
  \caption{The degree of saturation $S_r$ after the water supply stopped.}
  \label{saturation}
\end{figure}

The degrees of saturation $S_r$ can be calculated since we measured $W_0$ and $\phi$ (Fig.~\ref{saturation}). For aggregates, average of $D$ values defined as XS:~$D$ = 162~$\mu$m, S:~$D$ = 545 $\mu$m, M:~$D$ = 1480~$\mu$m, L:~$D$ = 3380~$\mu$m are used for the representative $D$ values. Regarding monomers, the samples were almost saturated when $D \leq$ 100 $\mu$m, while they were not fully saturated when $D \geq$ 400 $\mu$m. However, $S_r$ is kept large up to $D \sim$~2000 $\mu$m for aggregate soils. The large error of the data of monomer $D$ = 400 $\mu$m probably comes from the transitional behavior between the saturated regime and non-saturated regime.

\subsection{Comparison of monomer soil and aggregated soil}
Fig.~\ref{W0vsdD} displays the relation between $W_0$ and $d$ or $D$.
$W_0$ in aggregated soil was larger than that of non-aggregated (monomer) soil (Fig.~\ref{W0vsdD}(a)). Since the weight of glass beads forming soil sample was fixed, the number of particles composing the model soil is independent of the structure when $d$ is fixed. Thus, the total volume of microscopic pores should be almost identical when $d$ is identical (compare Fig.~\ref{structures}(a) bottom and (b)). However, volume of macroscopic pores is added in aggregated soil due to its hierarchical structure. Hence, the increase of $W_0$ in aggregated soil shown in Fig.~\ref{W0vsdD}(a) comes from the additional capacity of macroscopic pores which monomer soil does not possess.

In Fig.~\ref{W0vsdD}(b), the relation between $W_0$ and $D$ is presented.  Again, $W_0$ in aggregated soil is larger than that of monomer soil. In this case, however, the difference between monomer and aggregated soils of the same diameter $D$ is the existence of microscopic pores because aggregates can be regarded as a porous monomer particle (compare Fig.~\ref{structures}(a) upper and (b)). Since aggregates can retain water not only between the grains (macroscopic pores) but also inside the aggregates (microscopic pores), $W_0$ of the aggregated soil becomes larger than that of monomer soil.

\subsection{Characteristics in $d$-dependent and $D$-dependent behaviors of $W_0$}
\begin{figure}
  \centering
  \includegraphics[width=8cm]{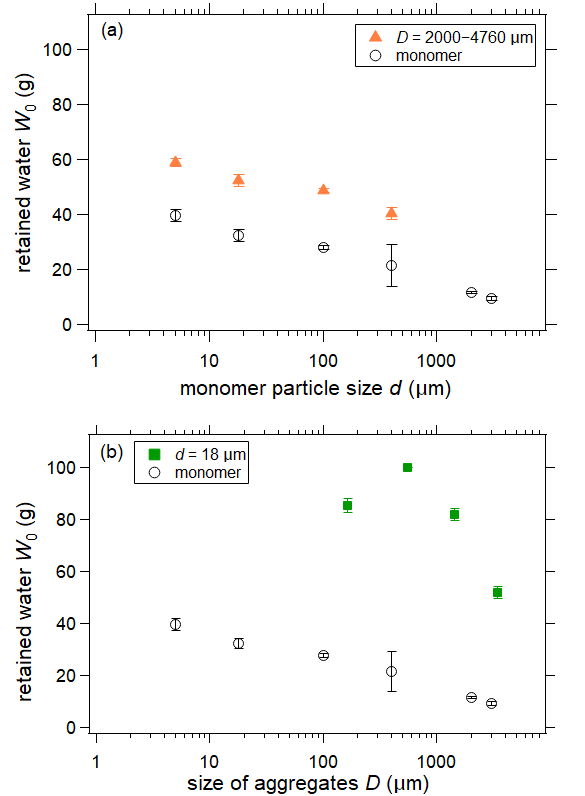}
  \caption{The relation between retained water $W_0$ in the steady state and (a) monomer particle size $d$ or (b) size of aggregates $D$. The monomer soil data plotted in both figures are identical because $d$ and $D$ cannot be distinguished in the case of non-aggregated (monomer) soil. }
  \label{W0vsdD}
\end{figure}

The negative correlation between $W_0$ and $d$ can be observed in both of monomer and aggregated soils (Fig.~\ref{W0vsdD}(a)). The water between grains is retained due to capillary force~\cite{Quere} via capillary bridges, in which the curvature radius is roughly proportional to $d$. Therefore, capillary-originated Laplace pressure decreases as $d$ increases. For large $d$ soils, gravity-driven drainage becomes dominant. The same tendency observed in aggregate soil can be explained by the same effect. Capillary effect is dominant in the microscopic pore scale because monomer size $d$ is considered. The amount of retained water is governed by the competition between capillary and drainage effects both in monomer and aggregated soils.

The most prominent feature confirmed in Fig.~\ref{W0vsdD}(b) is its non-monotonic behavior. Specifically, $W_0$ of aggregated soil shows the maximum around $D=500$~$\mu$m. All the aggregates used to obtain the data shown in Fig.~\ref{W0vsdD}(b) are composed of monomer glass beads of the constant diameter $d$ = 18 $\mu$m. Therefore, $W_0$ at $D$ = 18 $\mu$m must be identical to $W_0$ of monomer soil with $d$ = 18 $\mu$m. In other words, monomer soil and aggregated soil cannot be distinguished at $d$ = $D$. When aggregates of size $D$ ($>$ $d$) are formed, $W_0$ increases because an increase in $D$ creates larger macroscopic pores between aggregates. Thus, the hierarchical soil structure results in the increase in $W_0$. However, $W_0$ starts to decline in the range of gravity-dominant regime ($D \geq 500~\mu$m). 
Although this characteristic length scale $D\simeq 500$~$\mu$m is smaller than the typical capillary length of water ($\sim$ 2~mm)\cite{Quere}, we consider this value must be determined by the balance between capillary force and gravity force. Its specific value might be affected by the inhomogeneous sizes and shapes of the aggregates. For example, the aggregate soil plotted as $D$ = 545 $\mu$m is an average of the range of 250 - 840 $\mu$m and its shape is distorted (not spherical, see Fig.~\ref{sample}(c)).

\subsection{Discussion}
In the case of aggregated soil, $W_0$ is a sum of the retained water stored in microscopic pores $W_{\rm{micro}}$ and macroscopic pores $W_{\rm{macro}}$,
\begin{equation}
   W_0^{\rm{agg}}=W_{\rm{micro}}+W_{\rm{macro}},
\end{equation}
where $W_0^{\rm{agg}}$ is the total $W_0$ stored in the aggregated soil. 
When the monomer and aggregated soils are composed of same partcle size $d$, the balance between drainage and capillary suction can be assumed identical  in microscopic pores and independent of hierarchical structure. Besides, in this condition, the total volume of microscopic pores is also (roughly) identical because the mass of the model soil is fixed (100 g) (compare Fig.~1(a) bottom and (b)). Here, we can simply assume a relation,
\begin{equation}
W_{\rm{micro}}= W_0^{\rm{mono}},
\label{equal}
\end{equation}
where $W_0^{\rm{mono}}$ indicates $W_0$ of monomer soil. Thus, although it is difficult to distinguish $W_{\rm{micro}}$ and $W_{\rm{macro}}$ only from the experiment we conducted, $W_{\rm{macro}}$ can be estimated as, 
\begin{equation}
W_{\rm{macro}}=W_0^{\rm{agg}}-W_0^{\rm{mono}},
\end{equation}
with the assumption of Eq.~(\ref{equal}).

\begin{figure}
  \centering
  \includegraphics[width=8cm]{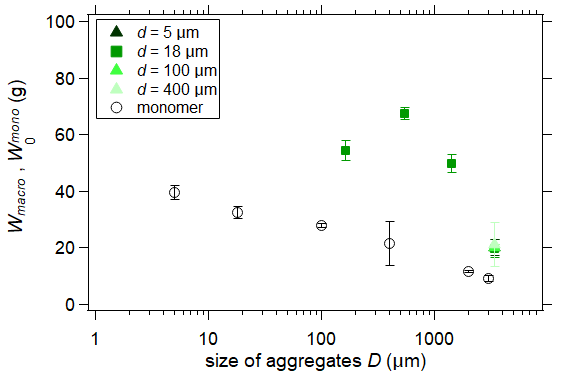}
  \caption{The relationship between the amount of retained water in macroscopic pores $W_{\rm{macro}}$ or $W_0^{\rm{mono}}$ and size of aggregates $D$.}
  \label{Wmacro}
\end{figure}

In Fig.~\ref{Wmacro}, $W_{\rm{macro}}$ and $W_0^{\rm{mono}}$ are compared in various $D$ cases. In this plot, the pores between monomer glass-beads are also regarded as "macroscopic" pores. As observed in Fig.~\ref{Wmacro}, $W_{\rm{macro}}$ of aggregated soil is approximately three times larger than $W_0^{\rm{mono}}$ around $D$ = 500~$\mu$m. This result indicates that water retention in macroscopic pores strongly depends on the hierarchical structure of the model soil. Due to the low $\phi$ of aggregated soil, the total volume of aggregated soil is greater than that of monomer soil in the fixed mass condition (compare Fig.~\ref{structures}(a) upper and (b)). This increased volume can be almost saturated at $D$ = 500 $\mu$m (Fig.~\ref{saturation}). However, $S_r$ gradually decreases in the range of $D >$ 500 $\mu$m.

	The variation of $S_r$ might result from the thickness of saturated aquifer supported by the capillary menisci among aggregates. To evaluate this effect, here we consider the balance between gravity and capillary effects. The former can be estimated by hydrostatic pressure and the latter can be modeled by Laplace pressure. Then, the balance can be written as,
\begin{equation}
  \frac{2\gamma}{R} \simeq \rho_w gH,
  \label{eq4}
\end{equation}
where $\gamma$ = 72.75 mN/m \cite{Quere}, $R$, $\rho_w$, $g$ and $H$ are surface tension, radius of the pore constriction, water density (1.0 $\mathrm{g/cm^3}$), gravitational acceleration (9.8 $\mathrm{m/s^2}$), and the thickness of the aquifer, respectively. High degree of saturation suggests that the retained water is connected to each other in the sample since the coalescence of capillary bridges results to liquid films across the porous material. Therefore, Laplace pressure at the bottom is an important element to retain water. The form of Eq.~(\ref{eq4}) actually corresponds to the definition of Bond number if we consider the characteristic length scales are $R$ and $H$, $B_0 = \rho_w gHR/\gamma$. This means that $H$ (water retention) is governed by the effective Bond number. The effective Bond number  should be in order of unity to satisfy the pressure balance. From the given values, $R$ must be less than 500--1000 $\mu$m in order to support the hydrostatic pressure with $H~\sim$ 1.5--3 cm which corresponds to the typical value of the sample thickness $H_{\rm{soil}}$. Steep decrease in $S_r$ in monomer soil from 100 $\mu$m to 400 $\mu$m (Fig.~\ref{saturation}) is roughly consistent with our estimation. The obtained $R$ value also agrees with aggregate diameter $D \simeq$ 500 $\mu$m at which $W_{\rm{macro}}$ shows a peak value. Gradual and later decrease in $S_r$ in aggregated soil compared to monomer soil is possibly due to the size distribution and shape anisotropy. Decreasing trend of $W_0^{\rm{mono}}$ and $W_{\rm{macro}}$ in the range of $D \geq$ 500 $\mu$m could be explained by the decrease in $H$. This effect suggests smaller $D$ is better able to support thick aquifer. However, $W_{\rm{macro}}$ is not a simple decreasing function. The volume of macroscopic pores depends on $D/d$. To secure the sufficient macroscopic pores, large $D$ is better. Too large and too small $D$ is not beneficial to retain water.

	$W_{\rm{macro}}$ is close to $W_0^{\rm{mono}}$ at $D$ $\simeq$ 3000~$\mu$m (L-sized aggregates) despite varied $d$. This indicates that the difference between monomer and aggregated soils is hardly found because the water drainage occurs in macroscopic pores in this range regardless of whether the grains are aggregates or monomers. 

Given the hydrophilicity of the glass beads, its wettability is needed to be taken into account in order to consider the application to natural soil situation since real soil has hydrophobic pockets/areas.  $W_0$ is expected to be smaller than the value we obtained when the soil is hydrophobic. However, we consider the tendency obtained in this study must be useful as a first-order approximation of the water retention characteristics in hierarchically structured soil. The water flow is certainly affected by grains wettability (e.g.~\cite{Wei:2014}). Thus, effect of grains wettability is an open future problem.

Direct observations of the retained water are crucial next step to further understand the water retention in the aggregated soil stored in macroscopic and microscopic pores. The state of water between grains (completely saturated or capillary bridge regime) could also be revealed by the direct observation such as X-ray microtomography~\cite{HERMINGHAUS1}. This will also enable us to reveal the complexity of hierarchical effects quantitatively and establish a more concise model.

\section{Conclusion}
The relation between the amount of retained water and two size parameters $d$ and $D$ characterizing aggregated soil structure was investigated in this study. From the measurement, the water drainage (dripping from the bottom sieve) was observed from the middle of spraying water and only right after stopping the water supply. To evaluate the ability of water retention in aggregated soil, we measured the amount of retained water in the steady state $W_0$ and analyzed its dependence on $d$ and $D$. As a consequence, we found some characteristic features of water retention in aggregated soil. First, $W_0$ in aggregated soil was larger than that of non-aggregated (monomer) soil. Second, $W_0$ decreased as $d$ increased when $D$ was fixed. Finally, we found $W_0$ showed the maximum value at $D\simeq 500$~$\mu$m when $d$ was fixed. Therefore, our result suggests that the smaller $d$ and $D\simeq 500$~$\mu$m are better to increase the water retention $W_0$. It is considered that the aggregated soil can efficiently retain water not only in microscopic pores within each aggregated particle but also between the aggregates around the size of $D\simeq 500$~$\mu$m. This specific value $D\simeq 500$~$\mu$m is determined by the balance between capillary force and gravity under the effect of complex geometry of aggregates and pore structure. In this study, only the spherical and hydrophilic glass beads were used to form soil. The effect of grains shape and their surface properties should be investigated to consider the application to the actual soil problem. In addition, internal water distribution should also be measured to fully understand the efficiency of water retention by macroscopic and microscopic pores. These are the important future problems.

\section*{Declaration of Competing Interest}
The authors declare that they have no known competing financial interests or personal relationships that could have appeared to influence the work reported in this paper.

\section*{Acknowledgment}
This work was supported by JSPS KAKENHI, Grant No. 18H03679.

\bibliographystyle{elsarticle-num} 
\bibliography{ref}

\begin{thebibliography}{10}
\bibitem{okayama}
Measurements of the mercury intrusion porosimetry of the samples were performed
  by Okayama Ceramics Center.
\end{thebibliography}

\end{document}